\documentclass[superscriptaddress,twocolumn,preprintnumbers,amssymb,nofootinbib]{revtex4-1}
\usepackage{amsmath, amsthm, amssymb} 

\usepackage{graphics,bm}
\usepackage{epsfig}
\usepackage{graphicx}
\usepackage{amsmath}
\usepackage{wrapfig}
\usepackage{color}
\usepackage{tikz}
\usetikzlibrary{arrows,shapes}
\usetikzlibrary{trees}
\usetikzlibrary{matrix,arrows} 			
\usetikzlibrary{positioning}				
\usetikzlibrary{calc,through}				
\usepackage{pgffor}							

\newcommand{\beq}{\begin{equation}}
\newcommand{\eeq}{\end{equation}}
\newcommand{\bqa}{\begin{eqnarray}}
\newcommand{\eqa}{\end{eqnarray}}

\newcommand{\ms}{\overline{\text{\tiny MS}}}

\def\square{\vcenter{\vbox{\hrule height.4pt
          \hbox{\vrule width.4pt height4pt
          \kern4pt\vrule width.3pt}\hrule height.4pt}}}

\begin{document}

\title{Bose-Einstein condensation and pion stars}

\author{Jens O. Andersen}
\email{andersen@tf.phys.ntnu.no}
\affiliation{Department of Physics, Faculty of Natural Sciences, NTNU, 
Norwegian University of Science and Technology, H{\o}gskoleringen 5,
N-7491 Trondheim, Norway}
\author{Patrick Kneschke}
\email{patrick.kneschke@uis.no}
\affiliation{Faculty of Science and Technology, University of Stavanger,
N-4036 Stavanger, Norway}
\date{\today}

\begin{abstract}
Pion stars consisting of Bose-Einstein condensed charged  pions
have recently been proposed as a new class of compact stars.
We use the two-particle irreducible effective action to leading
order in the $1/N$-expansion to describe
charged and neutrals
pions as well as the sigma particle. Tuning the parameters in the Lagrangian
correctly,
the onset of Bose-Einstein condesation of charged pions is exactly at
$\mu_I=m_{\pi}$, where $\mu_I$ is the isospin chemical potential.
We calculate the pressure, energy density, and equation of state, which
are used as input to the Tolman-Oppenheimer-Volkoff equations.
Solving these equations, we obtain the mass-radius relation for pion
stars. Global electric charge neutrality is ensured by adding
the contribution to the pressure and energy density from 
a gas of free relativistic leptons.
We compare our results with those of recent lattice simulations
and find good agreement. The masses of the pion stars are up to approximately
200 solar masses while the corresponding radii are of the order of
$10^5$ km.
\end{abstract}

\maketitle

\section{Introduction}
Apart from black holes, neutron stars are the most compact objects of
the universe. Their masses are 1-2 solar masses and their radii are 
of the order of 10km.
Ever since their existence was predicted by Landau
in 1932 \cite{landau}, have their properties been studied in detail.
One of the properties of interest is the mass-radius relation
of such compact objects. In order to obtain this relation, one
must solve the Tolman-Oppenheimer-Volkoff (TOV) equations, which
represent the 
generalization of hydrostatic equilibrium conditions in Newtonian gravity
to general relativity \cite{tov}. The TOV equations require
the equation of state (EoS) as input, i.e. one must know 
the EoS of nuclear matter at densities up to a few times saturation density.
It is well known that lattice Monte Carlo techniques cannot
be applied to systems with large baryon densities due to the infamous
sign problem. Consequently, the EoS and the properties of 
neutron stars can only be derived from model calculations, see
Ref. \cite{gordon} for a recent review.

Bose-Einstein condensation (BEC) occurs
in very different branches of physics ranging from condensation of
atoms in harmonic traps and condensation of $^4{\rm He}$ in 
superfluid Helium to condensation of pions and kaons in neutron stars
\cite{BEC}.
It is basically the phenomenon that a macroscopic number of bosons
occupy a specific single-particle state, which usually
is a zero-momentum state. 
In the context of QCD, the onset of pion condensation
at $T=0$ is when the 
isospin chemical potential $\mu_I$ is equal to the pion
mass,
$\mu_I^c=m_{\pi}$~\cite{isostep,isostep2}.~\footnote{
  Due to a another definition of
  the isospin chemical potential
  that differs by a factor of two,
  $\mu_I^c={1\over2}m_{\pi}$ is also frequently found in the
  literature.}
In two-flavor QCD with equal quark masses, there
is an $O(2)$ isospin symmetry, which gives
rise to a conserved isospin charge $Q_I$.
A pion condensate breaks this $O(2)$-symmetry 
and the phase transition from the
vacuum state to a Bose-condensed state is of second 
order for all temperatures.
\cite{kogut1,kogut2,gergy1,gergy2,gergy3}.
In contrast to QCD at finite baryon chemical
potential, there is no sign problem at finite isospin, and consequently
one can carry out lattice simulations. Some early results
can be found in Ref. \cite{kogut1,kogut2}, while recent results 
on the phase diagram in the $\mu_I-T$ plane are
reported in Ref. \cite{gergy1,gergy2,gergy3}. 

Various aspects of the QCD phase diagram at finite
isospin chemical potential have been studied using
chiral perturbation theory
(CHPT)~\cite{isostep,isostep2,split,loewe,fraga,cari},
the Nambu-Jona-Lasinio (NJL) model~\cite{2fbuballa,toublannjl,bar2f,he2f,heman2,heman,ebert1,ebert2,sun,lars,2fabuki,heman3,he3f}
and the quark-meson (QM) model
\cite{lorenz,ueda,qmstiele,chiralpion,patrick2} 
or their Polyakov-loop extended versions (PNJL and PQM).
The applicability
of Monte Carlo techniques offers the possibility of testing various
models directly. In Ref.~\cite{patrick2}, it was shown that all the
main features of the phase diagram mapped out in \cite{gergy1,gergy2,gergy3}
could be reproduced using the PQM model.
This includes the onset of charged pion condensation at $T=0$
at a critical isospin
chemical potential $\mu_I^c=m_{\pi}$, 
the second-order nature of this transition, and the merger of the
transition lines for the chiral transition and the BEC transition line, as well
as the BEC-BCS crossover at large values of $\mu_I$.

Boson stars have a long history since they were proposed almost 50 years
ago~\cite{bs1,bs2,bs3}. These stars are composed of
self-interacting bosons with e.g. a quartic interaction
term~\cite{colpi} and coupled to gauge fields~\cite{bij}.
Axion stars are a special type of boson stars involving
the hypothetical axion particle which originally was proposed to
solve the strong CP problem of QCD~\cite{kvin}. 
Axions are pseudo-Goldstone
bosons associated with the spontaneous breaking of a $U(1)$ symmetry
and form a Bose-Einstein condensate in these stars~\cite{eric}.

Recently, it has been proposed that
pions themselves may form a compact stellar object by condensing
into a zero-momentum state~\cite{endrodistar}.
Using the lattice results of Refs.~\cite{gergy1,gergy2,gergy3},
the authors of Ref. \cite{endrodistar}
calculated the EoS and the resulting
mass-radius relation with and without electric charge neutrality imposed.
Compared to a neutron star whose mass is of the order of one solar mass
and whose radius is of the order of 10 km, these new objects are huge;
their masses can be up to $M\approx200$ solar masses and their radii
as large as $1 0^5$ km \cite{endrodistar}.
In this paper, we will study pion stars using the two-particle
irreducible action formalism in the large-$N$ limit, where $N$
is the number of complex fields.
Setting $N=2$, this model reduces to the standard
$O(4)$-symmetric linear sigma model for the sigma particle and the three pions.

The article is organized as follows. In Sec. II, we discuss 
a self-interacting Bose gas in the context of the two-particle
irreducible (2PI) effective
action formalism and the $1/N$ expansion. 
We derive the pressure, energy density, and the pion condensate
as functions of the physical pion and sigma masses, the pion decay
constant and isospin chemical potential.
In Sec. III, we solve the Tolman-Oppenheimer-Volkoff equations
to obtain the mass-radius relation of a pion star with and without
electric charge neutrality.
In the Appendix, we briefly review the renormalization of the thermodynamic
functions and the matching of the parameters in the model in
the $\overline{\rm MS}$ scheme, and its relation to the parameters in the
on-shell scheme.
 
\section{Thermodynamics of an Interacting Bose gas}
We briefly discuss the application of the
2PI effective action formalism and the $1/N$-expansion of the
pressure, isospin density, and energy density.
Some details of the renormalization procedure can be 
found in the Appendix as well as in Refs. \cite{reinosa,jenson,hungary}.

The Euclidean Lagrangian for a Bose gas with $N$ species of
massive complex
scalars is 
\bqa\nonumber
{\cal L}&=&(\partial_{\mu}\Phi_i)^{\dagger}(\partial_{\mu}\Phi_i)
-{h\over\sqrt{2}}\left(\Phi_1^{\dagger}+\Phi_1\right)
+m^2\Phi_i^{\dagger}\Phi_i
\\ &&
+{\lambda\over2N}\left(\Phi_i^{\dagger}\Phi_i\right)^2\;,
\label{onlag}
\eqa
where $i=1,2,\dots,N$ and $\Phi_i={1\over\sqrt{2}}(\phi_{2i-1}+i\phi_{2i})$
are complex fields. If $h=0$, the symmetry of the Lagrangian (\ref{onlag})
is $O(2N)$, otherwise it is
$O(2N-1)$. For an $O(2N)$ symmetric theory, there are $(2N-1)N$
continuous symmetries
and each continuous symmetry gives rise to a conserved charged $Q_i$.
The maximum number of conserved charges that we can specify
simultaneously is the
maximum number of commuting generators, which is $N$, or $N-1$ if
$h\neq0$.
\cite{haber}.
Eventually we are 
interested in $N=2$ and the single chemical potential $\mu_I$
that in QCD corresponds to the conservation of isospin charge $Q_I$.
The chemical potential is introduced by replacing the partial
derivative with a covariant one, where $\mu_I$ is the zeroth component
of the gauge field. Identifying the complex field $\Phi_2$  
with the charged pions, the recipe is
$\partial_{\mu}\Phi_2\rightarrow(\partial_{\mu}-\mu_I\delta_{\mu}^0)\Phi_2$.
For $N=2$, the Lagrangian (\ref{onlag}) reduces to the linear sigma
model describing the three pions and the sigma particle.

In order to allow for a pion condensate $\rho_0$
in addition to a chiral condensate $\phi_0$,
we write the two complex fields $\Phi_1$ and $\Phi_2$
as
\bqa
\Phi_1&=&{1\over\sqrt{2}}\left(\phi_0+\phi_1+i\phi_2\right)\;,\\
\Phi_2&=&{1\over\sqrt{2}}\left(\rho_0+\phi_3+i\phi_4\right)\;.
\eqa
After symmetry breaking and with a nonzero pion condensate,
the thermodynamic potential $\Omega$ in the 2PI effective action formalism
can be written as 
\begin{widetext}
\bqa
\Omega&=&{1\over2}m^2(\phi_0^2+\rho_0^2)
+{\lambda\over8N}(\phi_0^2+\rho_0^2)^2
-{1\over2}\mu_I^2\rho_0^2-h\phi_0
+{1\over2}{\rm Tr}\ln D^{-1}+{1\over2}{\rm Tr}D_0^{-1}D
+\Phi[D]\;,
\label{omegadef}
\eqa
where $D$ is the exact propagator, $D_0$ is the tree-level propagator, and
$\Phi[D]$ is the sum of the two-particle irreducible diagrams.
The traces are over field indices as well as space-time.
The inverse tree-level propagator in Euclidean space can be written as 
\bqa
D_0^{-1}(P)&=&
\left(
\begin{array}{cccccc}
P^2+m_1^2&0&{\lambda\over N}\phi_0\rho_0&0&0&...\\
0&P^2+m_2^2&0
&0&0&...\\
{\lambda\over N}\phi_0\rho_0
&0&P^2+m_3^2&-\mu_IP_0&0&...\\
0&0&\mu_IP_0&P^2+m_4^2&0&...\\
0&0&0&0&P^2+m_2^2&...\\
\vdots&\vdots&\vdots&\vdots&\vdots&\ddots\\
\end{array}\right)\;,
\eqa
where $P^2={\bf P}^2+P_0^2$.
\end{widetext}
The tree-level masses are
\bqa
m_1^2&=&m^2+{3\lambda\over2N}\phi_0^2+{\lambda\over2N}\rho_0^2\;,
\\
m_2^2&=&m^2+{\lambda\over2N}\phi_0^2+{\lambda\over2N}\rho_0^2\;,
\\ 
m_3^2&=&-\mu_I^2+m^2+{\lambda\over2N}\phi_0^2+{3\lambda\over2N}\rho_0^2\;,\\
m_4^2&=&-\mu_I^2+m^2+{\lambda\over2N}\phi_0^2+{\lambda\over2N}\rho_0^2\;.
\eqa 
The terms in $\Phi[D]$ are $O(N)$ invariants and 
can be classified according to which 
order in the $1/N$-expansion they contribute. To leading order
the only contribution comes from
$\Phi_{\text{\tiny LO}}={\lambda\over8N}\left({\rm Tr}D\right)^2$,
which diagrammatically corresponds to
a double-bubble or figure-eight vacuum diagram.
The coupling gives a factor of $1/N$, while each trace yields a factor
of $N$.
The expectation values $\phi_0$ and $\rho_0$ satisfy the usual
stationarity conditions, while the exact propagator satisfies a
variational equation:
\bqa
{\delta\Omega\over\delta\phi_0}&=&0\;,
\hspace{1cm}
{\delta\Omega\over\delta\rho_0}=0\;,
\hspace{1cm}
{\delta\Omega\over\delta D}=0\;.
\label{gap}
\eqa 
Using that the self-energy $\Pi=D^{-1}-D_0^{-1}$, the variational 
gap equation can be written as $\Pi(D)=2{\delta\Phi\over\delta D}$.
At leading order we can write the inverse propagator of each particle
in the vacuum as 
$D_i^{-1}(P)=P^2+m_i^2+\Pi_{\text{\tiny LO}}(D)$.
The self-energy,
which is obtained by cutting a propagator line, corresponds to 
a tadpole diagram. The leading self-energy contribution
$\Pi_{\text{\tiny LO}}(D)$
is therefore a momentum-independent constant 
\bqa
\Pi_{\text{tniy LO}}&=&\lambda\int_Q{1\over Q^2+M^2}\;,
\eqa
where $M$ is a medium-dependent mass of the neutral pion
and the integral in Euclidean space is
\bqa
\int_Q&=&
\left({e^{\gamma_E}\Lambda^2\over4\pi}\right)^{\epsilon}
\int{d^dq\over(2\pi)^d}\;.
\eqa
Here $d=4-2\epsilon$ and $\Lambda$ is the renormalization scale
associated with dimensional regularization. 

\begin{widetext}
To leading order in the $1/N$-expansion, the three gap equations (\ref{gap}) are
\bqa
\label{gap1}
{\delta\Omega\over\delta\phi_0}&=&
m^2\phi_0+{\lambda\over2N}(\phi_0^2+\rho_0^2)\phi_0-h
+{\lambda\over2N}\phi_0{\rm Tr}D=
m^2\phi_0+{\lambda\over2N}(\phi_0^2+\rho_0^2)\phi_0-h+{\lambda}\phi_0
\int_Q{1\over Q^2+M^2}
=0\;,\\
{\delta\Omega\over\delta\rho_0}&=&
(m^2-\mu_I^2)\rho_0+{\lambda\over2N}(\phi_0^2+\rho_0^2)\rho_0+{\lambda}
\rho_0
\int_Q{1\over Q^2+M^2}=0\;,
\label{gap2}
\\
{\delta\Omega\over\delta D}&=&
{1\over2}{\rm Tr}\left(D^{-1}-D_0^{-1}\right)
+{\delta\Phi_{\text{\tiny LO}}\over\delta D}
=
M^2-m_2^2
-\lambda 
\int_Q{1\over Q^2+M^2}=0\;.
\label{gap3}
\eqa
\end{widetext}
These equations
are ultraviolet divergent and require nonperturbative
renormalization. The details of this procedure can be found
in Appendix A. At $T=0$, the system can be in two different phases
depending on the value of the isospin chemical potential $\mu_I$.
For $\mu_I\leq m_{\pi}$, the system is in the vacuum phase, where
$\phi_0=f_{\pi}$ and $\rho=0$. For $\mu_I > m_{\pi}$, the
system is in the pion-condensed phase, where $\phi_0={h\over\mu_I^2}$
and $\rho_0$ is nonzero.
This is shown below.
In the vacuum phase, one can identify
$M$ with the physical pion mass $m_{\pi}$, which follows from the fact
that the gap equation (\ref{gap3}) is the same as
the equation for the pole position of the pion mass (\ref{piv}).
In the pion-condensed phase,
we find $M=\mu_I$. This follows directly from subtracting
Eq. (\ref{gap2}) from Eq. (\ref{gap3}).
In the appendix, we discuss in some detail the renormalization of the
gap equation (\ref{gap3}). The other gap equations as well as the thermodynamic
potential $\Omega$ can be renormalized using the same techniques.

In the vacuum phase, the two nontrivial renormalized
gap equations (\ref{gap1}) and (\ref{gap3}) are
\bqa\nonumber
m_{\ms}^2\phi_0+{\lambda_{\ms}\over2N}\phi_0^3-h_{\ms}-
{\lambda\phi_0m_{\pi}^2\over(4\pi)^2}\left[\log{\Lambda^2\over m_{\pi}^2}+1\right]
&=&0\;,\\
\label{gap11}
\\ \nonumber
m_{\pi}^2-m_{\ms}^2-{\lambda_{\ms}\over2N}\phi_0^2
+{\lambda m_{\pi}^2\over(4\pi)^2}\left[\log{\Lambda^2\over m_{\pi}^2}+1\right]
&=&0\;,
\\
\label{gap33}
\eqa
where the $m^2_{\ms}$, $\lambda_{\ms}$, and $h_{\ms}$
are running parameters in the $\overline{\rm MS}$ renormalization
scheme
\bqa
m^2_{\ms}&=&{m^2_0\over1-{\lambda_0\over(4\pi)^2}
\log{\Lambda^2\over\Lambda_0^2}}\;,
\label{mrunning} \\
\lambda_{\ms}&=&{\lambda_0\over1-{\lambda_0\over(4\pi)^2}
  \log{\Lambda^2\over\Lambda_0^2}}\;,
\label{lrunning}\\
h_{\ms}&=&h\;,
\label{hrunning}
\eqa
and where $m_0^2$ and $\lambda_0$ are the values of the running parameters
at the scale $\Lambda_0$. In the appendix it is shown that 
if we choose $\Lambda^2=m_{\pi}^2/e$,
the parameters $m_0^2$ and $\lambda_0$ coincide with the parameters
$m_{\text{\tiny OS}}^2$ and $\lambda_{\text{\tiny OS}}$ in the on-shell scheme.
We note that $h$ does not require renormalization and consequently
$h_{\ms}=h=m_{\pi}^2f_{\pi}$ (see Appendix).

Combining the two equations (\ref{gap11})
and (\ref{gap33}), we see that $\phi_0$ satisfies
$m_{\pi}^2\phi_0=h=m_{\pi}^2f_{\pi}$
which implies that the minimum is at $\phi_0=f_{\pi}$
as it should.

In the pion-condensed phase, the renormalized gap equations are
\begin{widetext}
\bqa
\label{gap1p}
m_{\ms}^2\phi_0+{\lambda_{\ms}\over2N}\phi_0(\phi_0^2+\rho_0^2)
-h_{\ms}-
{\lambda_{\ms}
  \phi_0\mu_I^2\over(4\pi)^2}\left[\log{\Lambda^2\over\mu_I^2}+1\right]
&=&0\;,\\
(m_{\ms}^2-\mu_I^2)\rho_0+{\lambda_{\ms}\over2N}(\phi_0^2+\rho_0^2)\rho_0
-{\lambda_{\ms}\rho_0\mu_I^2\over(4\pi)^2}\left[\log{\Lambda^2\over\mu_I^2}+1\right]
&=&0\;,
\label{gap2p}
\\
\mu_I^2-m_{\ms}^2-{\lambda_{\ms}\over2N}(\phi_0^2+\rho_0^2)
+{\lambda_{\ms}\mu_I^2\over(4\pi)^2}\left[\log{\Lambda^2\over\mu_I^2}+1\right]
&=&0\;,
\label{gap3p}
\eqa
Combining Eqs. (\ref{gap1p}) and (\ref{gap3p}), we find
$\phi_0={h_{\ms}\over\mu_I^2}$. Using this result,
Eq. (\ref{gap2p}) can be written as
\bqa
\rho_0^2&=&{2N\over\lambda_{\ms}}\left(\mu_I^2-m^2_{\ms}\right)
-{h_{\ms}^2\over\mu_I^4}
+{2N\mu_I^2\over(4\pi)^2}\left[\log{\Lambda^2\over\mu_I^2}+1\right]
={(2\mu_I^2+m_{\sigma}^2-3m_{\pi}^2)f_{\pi}^2\over m_{\sigma}^2-m_{\pi}^2}
-{m_{\pi}^4f_{\pi}^2\over\mu_I^4}
+{2N\mu_I^2\over(4\pi)^2}\left[\log{m_{\pi}^2\over\mu_I^2}\right]
\;.
\label{roo}
\eqa
We are interested in the pressure in the two phases.
The pressure $P$ is given by minus the thermodynamic potential
$\Omega$ evaluated
at the solutions to the gap equations in the two phases.
Since we ultimately
want zero pressure in the vacuum phase, we calculate the pressure
difference of the two phases in the Appendix.
The result is
\bqa
P_{\text{eff}}&=&
{N\over2\lambda_{\ms}}\left(m_{\pi}^4-\mu_I^4\right)
+{1\over2}\mu_I^2\rho_0^2
+{m_{\pi}^4f_{\pi}^2\over\mu_I^2}-m_{\pi}^2f_{\pi}^2
-{N\mu_I^4\over2(4\pi)^2}\left[\log{\Lambda^2\over\mu_I^2}+{1\over2}\right]
+{Nm_{\pi}^4\over2(4\pi)^2}\left[\log{\Lambda^2\over m_{\pi}^2}+{1\over2}\right] 
\;.
\eqa
Using the running coupling constant Eq. (\ref{lrunning})
  and Eq. (\ref{roo}),
we obtain
\bqa
P_{\text{eff}}&=&
{1\over8}f_{\pi}^2{4m_{\sigma}^2(\mu_I^2-2m_{\pi}^2)+(2\mu_I^2-3m_{\pi}^2)^2+3m_{\pi}^4\over m_{\sigma}^2-m_{\pi}^2}
+{1\over2}{m_{\pi}^4f_{\pi}^2\over\mu_I^2}
+{N\mu_I^4\over2(4\pi)^2}\left[\log{m_{\pi}^2\over\mu_I^2}+{1\over2}\right]
-{Nm_{\pi}^4\over4(4\pi)^2}\;.
\label{pdiff}
\eqa
\end{widetext}
The pressure difference vanishes at threshold, $\mu_I=m_{\pi}$ as it should.
We can now make contact with tree-level results
in chiral perturbation theory as follows.
First, we ignore renormalization effects, i.e.
terms of order $N$ and then we 
simply take the limit $m_{\sigma}\rightarrow\infty$.
The results are
\bqa
\rho_0^{\text{\tiny CHPT}}&=&f_{\pi}\sqrt{1-{m_{\pi}^4\over\mu_I^4}}\;,\\
\label{ro}
P_{\text{eff}}^{\text{\tiny CHPT}}
&=&{1\over2}f_{\pi}^2\mu_I^2\left(1-{m_{\pi}^2\over\mu_{I}^2}\right)^2\;.
\label{pchpt}
\eqa
In the pion-condensed phase, $\phi_0={m_{\pi}^2f_{\pi}\over\mu_I^2}$
implying that $\phi_0^2+\rho_0^2=f_{\pi}^2$ independent of $\mu_I$.
One can therefore think of pion condensation as a rotation of
the chiral condensate into a pion condensate as the isospin
chemical potential increases.
\\
\indent
The pion-condensed phase is electrically charged, $n_Q\neq0$.
However, due to the Coulomb repulsion
among the pions, there is an enormous energy cost of having bulk matter
that is not electrically neutral~\cite{andreas}.
We will therefore impose electric
charge neutrality
and do so by adding
the free Lagrangian of a lepton field $l$ of mass $m_l$
\bqa
{\cal L}_{\rm lepton}&=&
\bar{l}\left[
i/\!\!\!\partial +\mu_l\gamma^0 +m_l\right]l\;,
\eqa
where $\mu_l$ is the lepton chemical potential.
This yields an extra contribution to the pressure, which at $T=0$ is given by
\bqa
\nonumber
P_l&=&2\int{d^3p\over(2\pi)^3} \left( \mu_l-\sqrt{p^2+m_l^2} \right)
\Theta\left(\mu_l-\sqrt{p^2+m_l^2}\right) \\ \nonumber
&=& {4\over3(4\pi)^2}\Big\{\mu_I\sqrt{\mu_l^2-m_l^2}
(\mu_l^2-\mbox{$5\over2$}m_l^2)
\\&&
+ {3\over2}m_l^4 \log{\mu_l + \sqrt{\mu_l^2-m_l^2}\over m_l}\Big\}
\Theta(\mu_l-m_l)\;.
\label{cont2}
\eqa
The total pressure
is then given by the sum of Eqs. (\ref{pdiff})
and (\ref{cont2}), and it is denoted by
$P(\mu_I,\mu_l)$.
The isospin and lepton number densities are given by
\bqa
\label{niso}
n_I(\mu_I) &=& {dP_{\text{eff}}\over d\mu_I} =
\mu_I\rho_0^2\;,
  \\ \nonumber
n_l(\mu_l) &=& {dP_l\over d\mu_l} 
    = {16\over3(4\pi)^2} \left(\mu_l^2-m_l^2\right)^{3\over2}\Theta(\mu_l-m_l)
\;,
\\
    \label{nlep}
\eqa
where the pion condensate is given by Eq.~(\ref{roo}).
Finally, the energy density is given
by 
\begin{widetext}
\bqa\nonumber
\epsilon(\mu_I,\mu_l) &=& -P(\mu_I,\mu_l) +\mu_I n_I(\mu_I) +\mu_l n_l(\mu_l)\\
\nonumber
&=&
    {1\over2}f_{\pi}^2
{m_{\sigma}^2(\mu_I^2+2m_{\pi}^2)+3\mu_I^2(\mu_I^2-m_{\pi}^2)-3m_{\pi}^4\over m_{\sigma}^2-m_{\pi}^2}
        -{3\over2}{m_{\pi}^4f_{\pi}^2\over\mu_I^2}+{3N\mu_I^4\over2(4\pi)^2}
        \log{m_{\pi}^2\over\mu_I^2}
        +{N(m_{\pi}^4-\mu_I^4)\over4(4\pi)^2}
        \\
        &&
        +
{4\over(4\pi)^2}\left[\mu_l\sqrt{\mu_l^2-m_l^2}(\mu_l^2-\mbox{$1\over2$}m_l^2)
- {1\over2}m_l^4 \log{\mu_l + \sqrt{\mu_l^2-m_l^2}\over m_l}\right]\;.
\eqa
\end{widetext}
The isospin density and energy density in CHPT follows from Eqs. (\ref{ro}),
(\ref{pchpt}) and (\ref{niso}).
We can write the latter in terms of the pressure and we finally obtain
\bqa
\label{ni}
n_I&=&\mu_If_{\pi}^2\left[1-{m_{\pi}^4\over\mu_I^4}\right]\;,
\hspace{1cm}\mu_I\geq m_{\pi}
\\
  {P\over\epsilon}&=&{\mu_I^2-m_{\pi}^2\over\mu_I^2+3m_{\pi}^3}\;,
\hspace{1.84cm}\mu_I\geq m_{\pi}\;.
\eqa
The onset of the isospin density is $\mu_I=m_{\pi}$ and it becomes
linear for large values of $\mu_I$. The ratio ${P\over\epsilon}$
vanishes at threshold, $\mu_I=m_{\pi}$ and approaches unity rather quickly
as $\mu_I$ increases.

\section{Numerical results and discussion}
In this section we will present and discuss our results.
The lepton is either the electron, $l=e$, or the muon, $l=\mu$.
The values for the meson and lepton
masses, and the pion-decay constant are 
\bqa
  m_\sigma &=& 600\;\text{MeV}\;,
  \hspace{1cm}
  m_\pi = 140\;\text{MeV}\;, \\
m_e &=& 0.511\;\text{MeV}\;,   \hspace{1cm}
m_\mu = 105\;\text{MeV}\;, \\
f_\pi &=& 93\;\text{MeV}\;.
\eqa
The sigma particle is a broad resonance, whose mass is in the 400-800 MeV range.
Unless otherwise stated, we will use the value $m_{\sigma}=600$ MeV which is a
fairly common choice. We will briefly discuss the $m_{\sigma}$
dependence of our results. 

\begin{figure}[htb]
\includegraphics[scale=0.9]{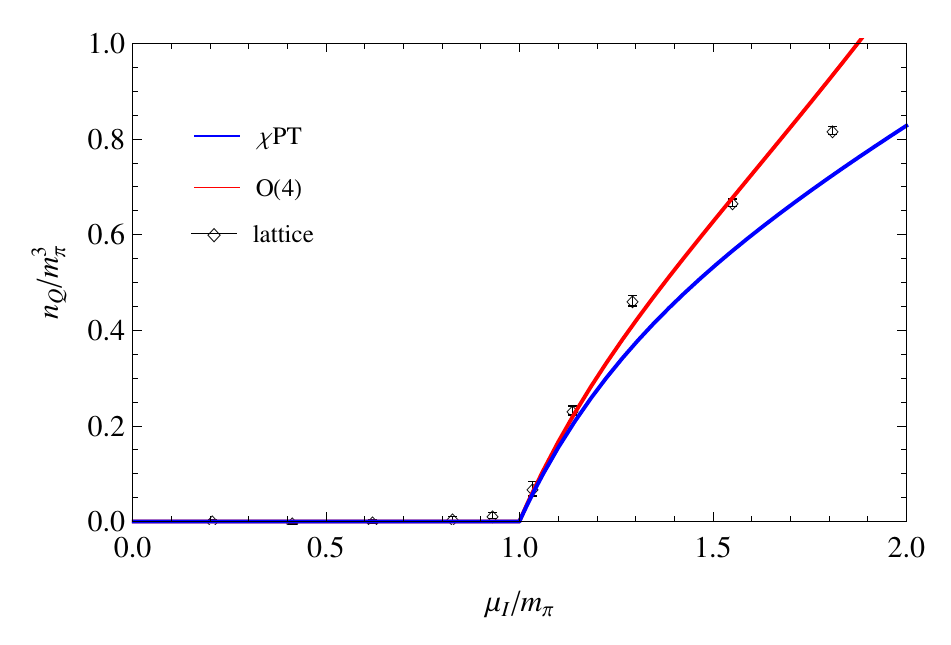}
\caption{Normalized electric charge density ${n_Q\over m_{\pi}^3}$
as a function of ${\mu_I\over m_{\pi}}$. See main text for details.}
\label{nqplot}
\end{figure}

In Fig.~\ref{nqplot}, we show the electric charge density $n_Q$ 
normalized by $m_{\pi}^3$ as a function of the isospin chemical potential
$\mu_I$ normalized by $m_{\pi}$.
The red line is Eq.~(\ref{niso})
and the blue line is from leading order
in chiral perturbation theory, Eq.~(\ref{ni}).
The data points are from the lattice simulations of
Brandt, Endrodi, and Schmalzbauer \cite{gergy1,gergy2,gergy3}
(The data points have been scaled since their defintion of $\mu_I$
differs by a factor of two compared to ours).
The charge density is zero all the way up to
$\mu_I=\mu_I^c=m_{\pi}$. This reflects the socalled Silver
Blaze property, which is the independence of physical quantities, such as the
isospin charge density, below some critical chemical potential. The
vacuum state of the theory is therefore defined by $\mu_I\leq\mu_I^c$.
The agreement between the results from the
lattice simulations and those from CHPT and the linear sigma model
is in general good, in particular for lower
values of $\mu_I$.

\begin{figure}[htb]
\includegraphics[scale=0.9]{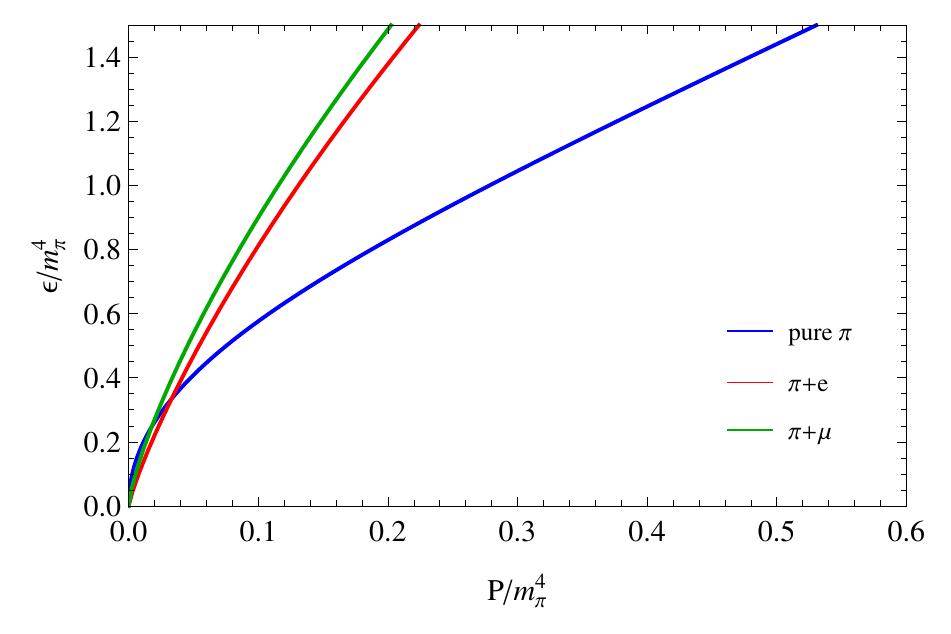}
\caption{Energy density ${\cal E}$ normalized to $m_{\pi}^4$ as a function
of the pressure normalized to $m_{\pi}^4$. See main text for details.}
\label{eos}
\end{figure}

In Fig. \ref{eos} we show the normalized equation of state, i.e. 
the energy density as a function of the pressure, both
normalized to $m_{\pi}^4$. The blue line is for a purely pionic system,
while the other lines are obtained when imposing charge neutrality by the
addition of muons (green) or electrons (red) to the system.
We notice that the imposition of electric charge neutrality
has a large effect on the EoS, although the mass dependence seems moderate
given the the two order of magnitude between the electron and muon masses.

Charge neutrality is given by the equation
\bqa
n_Q &=& {1\over2}n_I+n_l=0\;.
\label{nq}
\eqa
Inserting the isospin and lepton charge densities, given by Eqs. (\ref{niso})
and (\ref{nlep}), into Eq. (\ref{nq}) one obtains $\mu_I$ as a function
of $\mu_l$.

\begin{figure}[htb]
\includegraphics[scale=0.9]{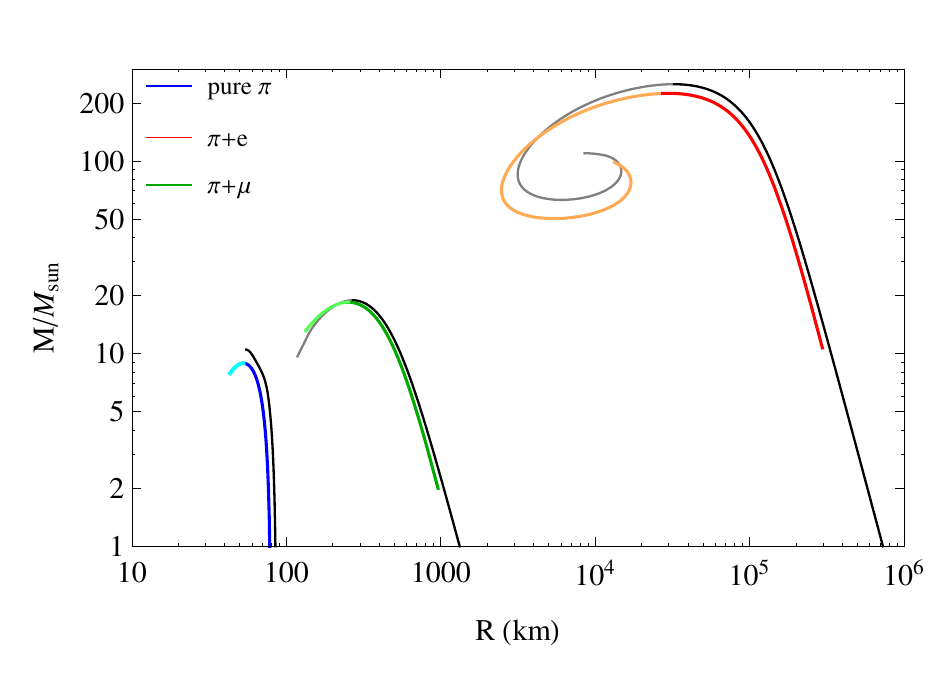}
\caption{Mass-radius relation of pion stars. See main text for details.}
\label{masr}
\end{figure}

We next determine the mass-radius relation of the pion star using the
Tolman-Oppenheimer-Volkoff equation. The radial dependence of the stellar mass
is given by the energy density
\bqa
{dm\over dr} = 4\pi^2r^2\epsilon(\mu_I,\mu_l)\;,
\eqa
while the TOV equation, which describes the pressure inside the star is
rewritten for the lepton chemical potential \cite{gergy3}
\bqa
{d\mu_l\over dr} = -G\mu_l\dfrac{m+4\pi r^3 P}{r^2-2Grm}
\left[1+2{\mu_I\over\mu_l}\right]\left[1+4{n_l^\prime\over n_I^\prime}\right]^{-1}
\;.
\label{tov2}
\eqa
For a pure pion star, the system is characterized by the isospin chemical
potential and Eq. (\ref{tov2}) reduces to
\bqa
{d\mu_I\over dr} &=& -G\mu_I\dfrac{m+4\pi r^3 P}{r^2-2Grm}\;.
\eqa
In Fig. \ref{masr}, we show the main result of the present paper, namely
the mass-radius relation of pion stars. 
The blue lines are for the pure pion system, where dark blue indicates the
stable solution of the TOV equation and light blue the unstable one. The dark
and light green lines are obtained by imposing charge neutrality by adding a
muon gas and the red and orange lines by adding electrons instead.
For comparison we
show the lattice results of \cite{gergy3} in black and grey, where we find
overall good agreement with our model.\\
The central isospin chemical potentials for the heaviest stable stars are
$\mu_I=252.88$ MeV (pure pion star), $\mu_I=142.0928$ MeV (pions+muons),
and $\mu_I=140.00008513$ MeV (pions+electrons).
The onset for pion condensation is $\mu_I^c=140$ MeV and therefore we find that
the pion pressure is very small in the pion-lepton systems, which results in a
larger star compared to the pure pion case.

\begin{figure}[htb]
\includegraphics[scale=0.6]{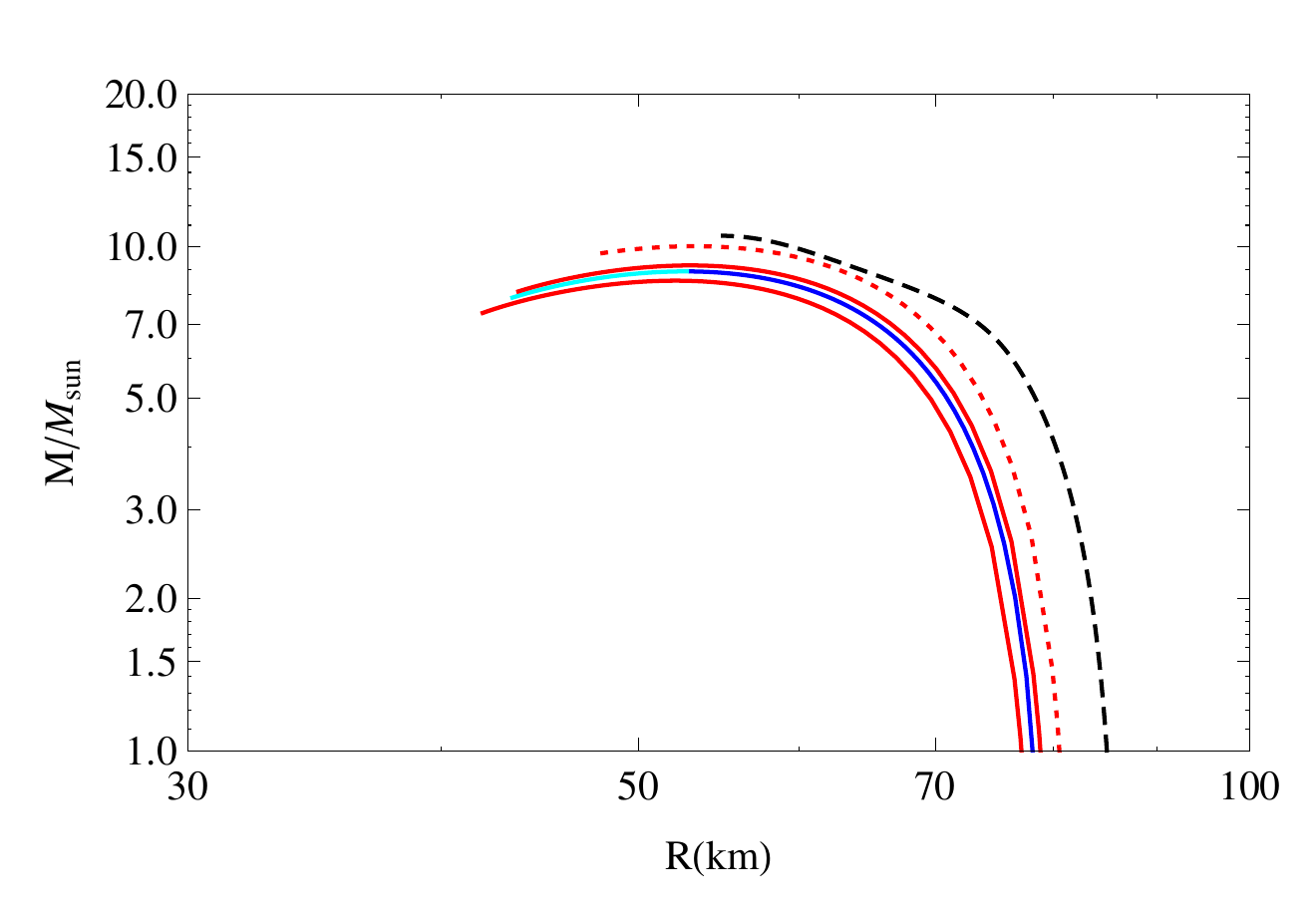}
\caption{Mass-radius relation of pion stars for different values of the
sigma mass. See main text for details.}
\label{masr2}
\end{figure}

In Fig. \ref{masr2}, we show the mass-radius relation for
pure pion stars and different values of the sigma mass.

The blue line is the result from Fig. \ref{masr}, i.e. for $m_{\sigma}=600$ MeV.
The solid red lines correspond to $m_\sigma=500$ MeV (lower) and $700$ MeV
(upper).
For comparison, the red dotted line shown the result from CHPT at tree level
and the black dashed line the  lattice results from Ref. \cite{gergy3}

Similar calculations for the neutral systems, with either muons or
electrons, show much smaller differences and are not shown here.
Based on the Fig. \ref{masr2} one might conclude that the result from 
CHPT, which is a model-independent one, is in best agreement with lattice
data. This is true for the pure pion star, which is the least interesting
case. It would be of interest to go to next-to-leading
order in chiral perturbation theory to study the convergence of the
results. Based on the values of the central isospin chemical potentials, 
one expects the largest corrections in the pure pion case.

\begin{figure}[htb]
\includegraphics[scale=0.9]{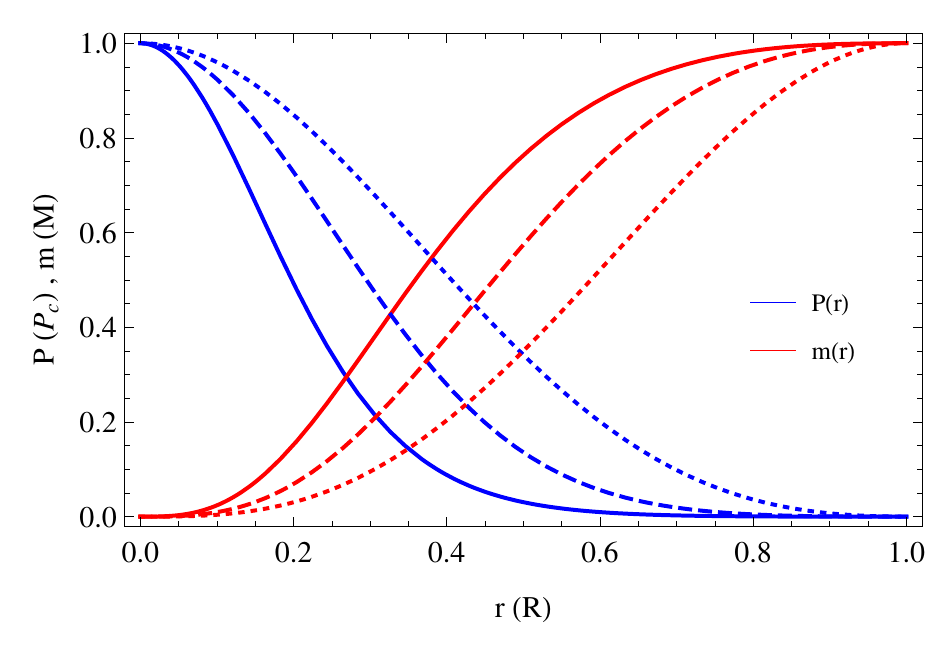}
\caption{Pressure and accumulated mass normalized to the central pressure
$P_c$ and the total mass $M$ of the pion star. See main text for details.}
\label{relations}
\end{figure}

In Fig. \ref{relations}, we show
the pressure $P(r)$ (blue lines) 
and accumulated mass $m(r)$ (red lines) normalized to the central pressure
$P_c$ and the total mass $M$ of the pion star, both as functions of the
normalized distance from the center. The solid lines are 
for $\pi$+$e$, dashed lines for $\pi$+$\mu$ and dotted lines for pions only.
The central pressure and mass correspond to the maximum of the $M(R)$ curves in
Fig. \ref{masr}. The faster the pressure decreases, the faster the mass
inside the star accumulates.
The typical central pressure we find for the charged pion star is of order
$P\sim 10^{33}\; \text{Pa}$. For the $\pi+\mu$ and $\pi+e$ system it is much
smaller, $P\sim 10^{31}\;\text{Pa}$ and $P\sim 10^{25}\;\text{Pa}$ respectively.
For comparison, the central pressure of neutron stars is
$P\sim 10^{34}\;\text{Pa}$.
Thus the higher the central pressure, the smaller the star.

\section*{Acknowledgements}
The authors would like to thank G. Endrodi and S. Schmalzbauer
for providing the
data points of the calculations in Ref. \cite{endrodistar}, as well
as useful discussions.

\appendix
\section{Renormalization of gap equations and parameter fixing}
In this Appendix, we carry out renormalization of the nonperturbative
gap equations. We also briefly discuss how one can use the
$\overline{\rm MS}$ and the
on-shell renormalization schemes to express the running
parameters in terms of meson masses, the pion decay constant, and the
renormalization scale.

In order to renormalize the gap equations and pressure, 
we need the following divergent integrals in dimensional regularization for
$d=4-2\epsilon$ dimensions
\bqa\nonumber 
C(m^2)&=&\int_Q\log\left[Q^2+m^2\right]
\\&=&-{m^4\over2(4\pi)^2}\left({\Lambda^2\over m^2}\right)^{\epsilon}
\left[{1\over\epsilon}+{3\over2}+{\cal O}(\epsilon)\right]\;,
\label{cdef}
\\
\nonumber
A(m^2)&=&\int_Q{1\over Q^2+m^2}
\\&=&-{m^2\over(4\pi)^2}\left({\Lambda^2\over m^2}\right)^{\epsilon}
\left[{1\over\epsilon}+1+{\cal O}(\epsilon)\right]\;,
\label{adef}
\eqa
where $\Lambda$ is the renormalization scale associated with dimensional
regularization.

We first renormalize the gap equation (\ref{gap3}), 
which we rewrite as
\bqa
M^2-m_2^2&=&
\lambda\int_Q{1\over Q^2+m_2^2+\Pi_{\text{\tiny LO}}}\;,
\eqa
where we have used that
$M^2=m^2_2+\Pi_{\text{\tiny LO}}$.
Expanding the right-hand side in powers
of $\Pi_{\text{\tiny LO}}$, we find
\bqa\nonumber
{M^2\over\lambda}-{m_2^2\over \lambda}&=&
\int_Q{1\over Q^2+m_2^2}
-\Pi_{\text{\tiny LO}}\int_Q{1\over(Q^2+m_2^2)^2}\\
&&
  +  \Pi_{\text{\tiny LO}}^2\int_Q{1\over(Q^2+m_2^2)^3}
+...
\label{gapleik}
\;.
\eqa
The self-energy is written in a power series in $\lambda$,
$\Pi_{\text{\tiny LO}}=\Pi_{\text{\tiny LO}}^{(1)}+\Pi_{\text{\tiny LO}}^{(2)}+...$
where the superscript
indicates the power of $\lambda$.
Introducing renormalization constants $\delta m_i^2$ and $\delta \lambda_i$ for
each order in the coupling, Eq. (\ref{gapleik}) can be renormalized
iteratively. The first iteration gives
\bqa
\Pi_{\text{\tiny LO}}^{(1)}=\lambda\int_Q{1\over Q^2+m_2^2}\;,
\eqa
and using
Eq. (\ref{adef}) we find
\bqa\nonumber
{M^2}&=&{m^2+\delta m_1^2}+{\lambda+\delta\lambda_1\over2N}
\left(\phi_0^2+\rho_0^2\right)
\\ &&
-{\lambda m_2^2\over(4\pi)^2}\left[{1\over\epsilon}+\log{\Lambda^2\over m_2^2}+1
  \right]\;,
\label{lodiv}
\eqa
including the leading-order counterterms
$\delta m_1^2$ and $\delta\lambda_1$.
The divergences in (\ref{lodiv}) are removed by choosing the counterterms
in the $\overline{\rm MS}$ scheme
\bqa
\delta m^2_1&=&{\lambda m^2\over(4\pi)^2\epsilon}
\;,\hspace{1cm}
\delta\lambda_1={\lambda^2\over(4\pi)^2\epsilon}\;.
\eqa
Carrying out renormalization order by order in $\lambda$, one finds that the
$n$'th order mass and coupling constant counterterms
are~\cite{reinosa,jenson}
\bqa
\delta m_n^2&=&{\lambda^n m^2\over(4\pi)^{2n}\epsilon^n}
\;,\hspace{1cm}
\delta\lambda_n={\lambda^{n+1}\over(4\pi)^{2n}\epsilon^n}\;.
\label{deltan}
\eqa
The coupling $\lambda$ is renormalized by writing
$\lambda \rightarrow \lambda_{\rm b}=\Lambda^{2\epsilon}(\lambda_{\rm \ms}+\delta\lambda)$, where $\lambda_{\rm b}$ is the bare coupling and 
\bqa
\delta\lambda&=&\sum_{n=1}^{\infty}\delta\lambda_n
={\lambda^2\over(4\pi)^2\epsilon}{1\over1-{\lambda\over(4\pi)^2\epsilon}}\;.
\eqa
Solving for ${1\over\lambda_{\rm b}}$ yields
\bqa
      {1\over\lambda_{\rm b}} 
      &=&
      {\Lambda^{-2\epsilon}\over\lambda_{\ms}}-{\Lambda^{-2\epsilon}\over(4\pi)^2\epsilon}
      \;.
\label{msla}
\eqa
It follows from Eq. (\ref{deltan}) that
$\delta m^2={m^2_{}\over\lambda_{}}\delta\lambda$ and therefore
\bqa
{m^2_{\rm b}\over\lambda_{\rm b}}={m^2_{\ms}\over\lambda_{\ms}}.
\label{msm}
\eqa
Let us rewrite the gap equation (\ref{gap3}) as
\bqa\nonumber
    {M^2\over\lambda}&=&{m^2\over\lambda}+{\phi_0^2+\rho_0^2\over2N}
    +\int_Q{1\over Q^2+M^2}
 \\ \nonumber
 &=&
{m^2\over\lambda}+{\phi_0^2+\rho_0^2\over2N}
-{M^2\over(4\pi)^2}\left[
{1\over\epsilon}+\log{\Lambda^2\over M^2}+1
\right]\;.
\\
\label{gap44}
\eqa
The gap equation (\ref{gap44}) 
is made finite by using Eq. (\ref{msm}) and 
substituting Eq. (\ref{msla}). The result is
\bqa\nonumber
M^2&=&m_{\ms}^2+{\lambda_{\ms}(\phi_0^2+\rho_0^2)\over2N}
-{\lambda_{\ms}M^2\over(4\pi)^2}\left[
\log{\Lambda^2\over M^2}+1
\right]\;,
\\
\eqa
which is Eq. (\ref{gap33}) in the vacuum phase and Eq. (\ref{gap3p})
in the BEC phase.
The gap equation (\ref{gap1})
can be renormalized in the same manner and the result is given in
Eq. (\ref{gap11}) and (\ref{gap1p}), respectively. The
renormalized version of Eq. (\ref{gap2}) for the pion condensate
is (\ref{gap2p}).
Combining Eq. (\ref{gap11}) and (\ref{gap33}),
we find $h=m_{\pi}^2f_{\pi}$, i.e. the tree-level relation.
Thus the parameter $h$ is not renormalized. 

Taking the derivative of Eq. (\ref{msla}) with respect to the renormalization
scale $\Lambda$ and using that the bare coupling $\lambda_b$
is independent 
of $\Lambda$, we find that the renormalized coupling $\lambda_{\ms}$ satisfies
a renormalization group equation. In the limit $\epsilon\rightarrow0$,
this equation reads
\bqa
      \Lambda{d\lambda_{\ms}\over d\Lambda}&=&
          {2\lambda_{\ms}^2\over(4\pi)^2}\;,
\label{rg1}
\eqa
whose solution is given by
\bqa
\lambda_{\ms}&=&
{\lambda_0\over1-{\lambda_0\over(4\pi)^2}\log{\Lambda^2\over\Lambda_0^2}}\;,
\label{lrun}
\eqa
where the constant $\lambda_0$ is the value of the running
parameter $\lambda_{\ms}$ at the scale $\Lambda_0$.
Using ${m^2_{\rm b}\over\lambda_{\rm b}}={m_{\ms}^2\over\lambda_{\ms}}$ and that the
bare mass and coupling are independent of the renormalization scale,
as well as Eq. (\ref{rg1}), one can show that the renormalized 
mass similarly satisfies the 
renormalization group equation,
\bqa
\Lambda{dm^2_{\ms}\over d\Lambda}&=&
{2\lambda_{\ms}m_{\ms}^2\over(4\pi)^2}\;.
\label{rg2}
\eqa
The solution to Eq. (\ref{rg2}) is
\bqa
m^2_{\ms}&=&
{m^2_0\over1-{\lambda_0\over(4\pi)^2}\log{\Lambda^2\over\Lambda_0^2}}\;,
\label{mrun}
\eqa
where the constant $m^2_0$ is the value of the running mass
parameter $m_{\ms}^2$ at the scale $\Lambda_0$.

As mentioned in the main text we are ultimately interested in the
pressure. We first write 
${1\over2}m^2(\phi_0^2+\rho_0^2)+{\lambda\over8N}(\phi_0^2+\rho_0^2)^2$ as
${N\over2\lambda}[m^2+{\lambda\over2N}(\phi_0^2+\rho_0^2)]^2-{Nm^4\over2\lambda}$
in the potential Eq.~(\ref{omegadef}) and then
use the gap equation (\ref{gap3}).
The resulting unrenormalized pressure reads
\bqa\nonumber
P&=&{N\over2\lambda}\left(m^4-M^4\right)+{1\over2}\mu_I^2\rho_0^2
+h\phi_0\\&& \nonumber
-N\int_Q\log\left[Q^2+M^2\right]
+NM^2\int_Q{1\over Q^2+M^2}\;.
\\ &&
\eqa
The pressure difference of the two phases is then
\bqa\nonumber
P_{\text{eff}}&=&
{N\over2\lambda}\left(m_{\pi}^4-\mu_I^4\right)+{1\over2}\mu_I^2\rho_0^2
+{m_{\pi}^4f_{\pi}^2\over\mu_I^2}-m_{\pi}^2f_{\pi}^2
\\ \nonumber
&&-N\int_Q\log\left[Q^2+\mu_I^2\right]
+N\mu_I^2\int_Q{1\over Q^2+\mu_I^2}\;.
\\ \nonumber
&&+N\int_Q\log\left[Q^2+m_{\pi}^2\right]
-Nm_{\pi}^2\int_Q{1\over Q^2+m_{\pi}^2}\;.
\\
\eqa
Using the integrals (\ref{cdef})--(\ref{adef}) and renormalizing the coupling
according to Eq. (\ref{msla}),
the renormalized pressure difference is
\bqa\nonumber
P_{\text{eff}}&=&
{N\over2\lambda_{\ms}}\left(m_{\pi}^4-\mu_I^4\right)+{1\over2}\mu_I^2\rho_0^2
+{m_{\pi}^4f_{\pi}^2\over\mu_I^2}-m_{\pi}^2f_{\pi}^2
\\ \nonumber
&&-{N\mu_I^4\over2(4\pi)^2}\left[\log{\Lambda^2\over\mu_I^2}+{1\over2}\right]
\\
&&
+{Nm_{\pi}^4\over2(4\pi)^2}
\left[\log{\Lambda^2\over m_{\pi}^2}+{1\over2}\right]\;.
\eqa
We finally discuss renormalization in the on-shell
scheme~\cite{sir1,sir2,hollik}.
The counterterms in this scheme
are determined by demanding that the renormalized
mass is equal to the physical mass, i.e. the pole mass, and
that the residue of the propagator is unity,
\bqa
\label{os1}
\Pi_{\sigma,\pi}(P^2=m_{\sigma,\pi}^2)+\text{counterterms}&=&0\;,\\
{\partial\over\partial P^2}\Pi_{\sigma,\pi}(P^2)\big|_{P^2=m_{\sigma,\pi}^2}
+\text{counterterms}
&=&0\;,
\label{os2}
\eqa
where $\Pi_{\sigma,\pi}(P^2)$ is the self-energy function.
Eq. (\ref{os2}) is trivially satisfied in the present case since
the leading order self-energy is independent of the external momentum.
The inverse propagator for the sigma and pion can be written as
\bqa
\label{prop1}
P^2+m_1^2+\delta m_1^2+\Pi_{\text{\tiny LO}}\;, \\
    P^2+m_2^2+\delta m_2^2+\Pi_{\text{\tiny LO}}\;.
\label{prop2}
\eqa
Evaluating Eqs. (\ref{prop1}) and (\ref{prop2}) on-shell and 
suppressing the counterterms, the 
equations for the pion and sigma
masses in the vacuum then become
\bqa
\label{piv}
m_{\pi}^2&=&m^2+{\lambda\over2N}f_{\pi}^2+
\lambda\int_Q{1\over Q^2+m_{\pi}^2}\;,\\
\label{sigmav}
m_{\sigma}^2&=&m^2+{3\lambda\over2N}f_{\pi}^2+
\lambda\int_Q{1\over Q^2+m_{\pi}^2}\;.
\eqa
We note in passing that Eq. (\ref{piv}) is identical to Eq. (\ref{gap3})
in the vacuum phase.
At tree level, we can express the parameters $m^2$, $\lambda$, and $h$
in terms of the physical sigma and pion masses, as well as the
pion decay constant,
\bqa
\label{mpar}
m^2&=&-{1\over2}(m_{\sigma}^2-3m_{\pi}^2)\;,\\
\lambda&=&N{(m_{\sigma}^2-m_{\pi}^2)\over f_{\pi}^2}\;,
\label{lpar}\\
h&=&m_{\pi}^2f_{\pi}\;.
\label{hbar}
\eqa
We first consider Eq. (\ref{piv}).
Again we rewrite it as
\bqa\nonumber
{m_{\pi}^2\over\lambda}&=&
{m^2\over\lambda}+{f_{\pi}^2\over2N} +\int_Q{1\over Q^2+M^2}
 \\
 &=&
{m^2\over\lambda}-{f_{\pi}^2\over2N}
-{m_{\pi}^2\over(4\pi)^2}\left[
{1\over\epsilon}+\log{\Lambda^2\over m_{\pi}^2}+1
\right]\;.
\label{vacuumpi}
\eqa
The self-energy correction $\Pi_{\text{\tiny LO}}$ can be eliminated by
the renormalization of the coupling,
$\lambda_b= \Lambda^{2\epsilon} ( \lambda_{\text{os}} +\delta\lambda )$, with
\bqa
    {1\over\lambda_{\rm b}}
    &=&
 {\Lambda^{-2\epsilon}\over\lambda_{\text{os}}}-{\Lambda^{-2\epsilon}\over(4\pi)^2}
      \left[{1\over\epsilon}+\log{\Lambda^2\over m_{\pi}^2}+1\right]\:.
      \label{osla}
      \eqa
Furthermore, if we define $\delta m^2=m^2\delta\lambda$, i.e.
${m_{\text{b}}^2\over\lambda_{\text{b}}}={m^2_{\text{os}}\over\lambda_{\text{os}}}={m^2\over\lambda}$, Eq. (\ref{vacuumpi})
reduces to the tree-level expression, as it should in the OS-scheme.
That this recipe is consistent
can be seen by renormalizing Eq. (\ref{vacuumpi}) iteratively
as we did above. Then one finds
\bqa
\label{deltan22}
\delta m_n^2&=&{\lambda^{n}\over(4\pi)^{2n}} m^2
\left[{1\over\epsilon}
+\log{\Lambda^2\over m_{\pi}^2}+1\right]^n\;, \\
\delta\lambda_n&=&{\lambda^{n+1}\over(4\pi)^{2n}}\left[{1\over\epsilon}
    +\log{\Lambda^2\over m_{\pi}^2}+1
\right]^n
    \;.
\label{deltan2}
\eqa
It is straightforward to show that $\lambda_{\text{b}}$ and
  $m_{\text{b}}^2$
are independent of the renormalization scale $\Lambda$ as they must be.
  Moreover, using the fact that the bare parameters are independent of
  the renormalization scheme we can use Eqs. (\ref{msla}) and (\ref{osla}) to
  show that $\lambda_0 = \lambda$, given by Eq. (\ref{lpar}) if we choose
  $\Lambda_0^2={m_{\pi}^2\over e}$.
From Eqs. (\ref{lrun}) and (\ref{mrun}) it then follows that
$m_0^2$ is given by the right-hand side of
Eqs. (\ref{mpar}).
After having renormalized the gap equation defining the
pion mass nonperturbatively, we make some remarks regarding
the equation for the sigma particle. While the 
divergence in Eq. (\ref{sigmav}) is the same as in (\ref{piv}), the
tree-level term involving the coupling is three times as large.
Thus it seems that one cannot renormalize the gap equation for the
sigma mass using the counterterms given by Eqs.
(\ref{deltan22})--(\ref{deltan2}).
The solution to this problem is to realize that one  must include
all counterterms that respect the symmetries. In the present case there
is a counterterm proportional to ${\rm Tr}(G^2)$ which is of order
one, i.e. next-to-leading in the $1/N$-expansion \cite{hungary}.
One can then write the coupling constant counterterm in Eq. (\ref{sigmav})
as~\cite{hungary}
\bqa
{\delta\lambda_A+2\delta\lambda_B\over6N}f_{\pi}^2\;,
\eqa
where the two terms
contribute at order $N$ and one, respectively.
The term $\delta\lambda_A$ is equal to the counterterm we have already found,
while the term $\delta\lambda_B$ is used for renormalizing the gap equation
at next-to-leading order.
Finally, we remark that the parameter $h$ 
does not require renormalization
and therefore $h_{\ms}=h_{\text{\rm os}}=m_{\pi}^2f_{\pi}$.


\end{document}